\documentclass[10pt, journal]{IEEEtran}

\usepackage{multirow}
\usepackage{booktabs}
\usepackage{color}
\usepackage{graphicx}
\usepackage[tight,footnotesize]{subfigure}
\usepackage{stfloats}
\usepackage{amsmath,amsthm,amssymb,amsfonts}
\usepackage{bm}
 \usepackage{subeqnarray}
 \usepackage{fancyhdr}
\usepackage{url}
\usepackage[justification=justified]{caption}
\ifCLASSOPTIONcompsoc
  \usepackage[nocompress]{cite}
\else
  \usepackage{cite}
\fi
\usepackage{geometry}
\begin{document}

\title{Spatio-Temporal Representation with Deep Neural \\Recurrent Network in MIMO CSI Feedback}
%

\markboth{IEEE Wireless Communications Letters,~Vol.~, No.~, ~2019}%
{Li \MakeLowercase{\textit{et al.}}: Spatio-Temporal Representation with Deep Recurrent Neural Network in MIMO CSI Feedback}


\author{Xiangyi~Li and~Huaming~Wu,~\IEEEmembership{Member,~IEEE}

\IEEEcompsocitemizethanks{
\IEEEcompsocthanksitem Y. Li and H. Wu are with the Center for Applied Mathematics, Tianjin University, Tianjin 300072, China (e-mail: xiangyi\_li@tju.edu.cn; whming@tju.edu.cn).
\IEEEcompsocthanksitem This work was supported by the National Key R $\&$ D Program of China (2018YFC0809800), the National Natural Science Foundation of China (61801325), and the Huawei Innovation Research Program (HO2018085138). (Corresponding author: Huaming Wu.)}
 }

\maketitle

\begin{abstract}
In multiple-input multiple-output (MIMO) systems, it is crucial of utilizing the available channel state information (CSI) at the transmitter for precoding to improve the performance of frequency division duplex (FDD) networks. One of the main challenges is to compress a large amount of CSI in CSI feedback transmission in massive MIMO systems. In this paper, we propose a  deep learning (DL)-based approach that uses a deep recurrent neural network (RNN) to learn temporal correlation and adopts depthwise separable convolution to shrink the model. The feature extraction module is also elaborately devised by studying decoupled spatio-temporal feature representations in different structures. Experimental results demonstrate that the proposed approach outperforms existing DL-based methods in terms of recovery quality and accuracy, which can also achieve remarkable robustness at low compression ratio (CR).


\end{abstract}


\IEEEpeerreviewmaketitle

\section{Introduction}\label{sec:introduction}
\IEEEPARstart{T}{he} technology of massive multiple-input multiple-output (MIMO), which was first pointed out in the early twentieth century, has become increasingly crucial in new generation mobile wireless communications (5G or B5G). The system uses multiple antennas as multiple transmitters at the base station (BS) and receivers at user equipment (UE) to realize the multipath transmitting, which can double the channel capacity without increasing spectrum resources or antenna transmit power. A growing number of studies \cite{o2017deep, wang2018deep, ref4 } have shown the significance of utilizing the channel state information (CSI) feedback at the transmitter to gain the improvement of MIMO systems. In a frequency division duplex (FDD) network \cite{choi2014downlink}, UE can estimate the downlink CSI, which is then fed back to the BS to perform precoding for the next signal.

 In fact, the uplink CSI feedback process is not an easy task in massive MIMO systems \cite{ref3}, due to a large number of antennas at the BS, resulting in high CSI feedback and huge computational complexity. In order to reduce the CSI feedback overhead, many methods and technologies have been proposed. Some compressive sensing (CS)-based approaches may not fit in real world CSI feedback systems and perform poorly in CSI compression due to the harsh preconditions. Recent studies have shown that applying DL to address the nonlinear problems or challenges in wireless communications can boost the quality of CSI feedback compression \cite{ref4}. Wen \textit{et al}. \cite{ref9} proposed an autoencoder network called CsiNet, which used several neural network (NN) layers as an encoder instead of the CS model to compress CSI as well as a decoder to recover the original CSI. Furthermore, they also put forward another network called CsiNet-LSTM \cite{wang2018deep}, which extended CsiNet with three RNN layers to show the benefits of exploring temporal channel correlation. Another paralleled work, called RecCsiNet \cite{lu2018mimo}, applied RNN in both the encoder and decoder to reduce errors in CSI compression and decompression. Both of them can improve the performance of the CsiNet network to some extent and outperform state-of-the-art CS methods.

 In this paper, we design a new architecture of deep NN in CSI feedback compression, which also takes advantage of RNN. Based on the RecCsiNet architecture, we retain its structure of feature compression and decompression modules, and further improve the feature extraction by applying RNN and separating feature extraction in the spatial and temporal domains. In addition, motivated by MobileNet \cite{sandler2018mobilenetv2} that used depthwise separable convolutions to build lightweight deep NN, which we substitute them for standard convolutions to enhance the quality of RefineNet \cite{ref9}. The main contributions are summarized as follows:
  \begin{itemize}
  \item We propose a novel and effective CSI sensing and recovery mechanism in the FDD MIMO system, referred to as ConvlstmCsiNet, which takes advantage of the memory characteristic of RNN in modules of feature extraction, compression and decompression, respectively. Moreover, we adopt depthwise separable convolutions in feature recovery to reduce the size of the model and interact information between channels.

  \item We further refine ConvlstmCsiNet in the feature extraction module by exploring the spatial-temporal feature representation that decouples a convolution in the spatial and temporal domains. Experimental results demonstrate that the improved ConvlstmCsiNet achieves the highest recovery quality at different compression ratios (CRs) compared to the state-of-the-art DL-based models.

\end{itemize}

\begin{figure*}
  \centering
 \includegraphics[width=5.8in, trim=0in 0in 0in 0.3in]{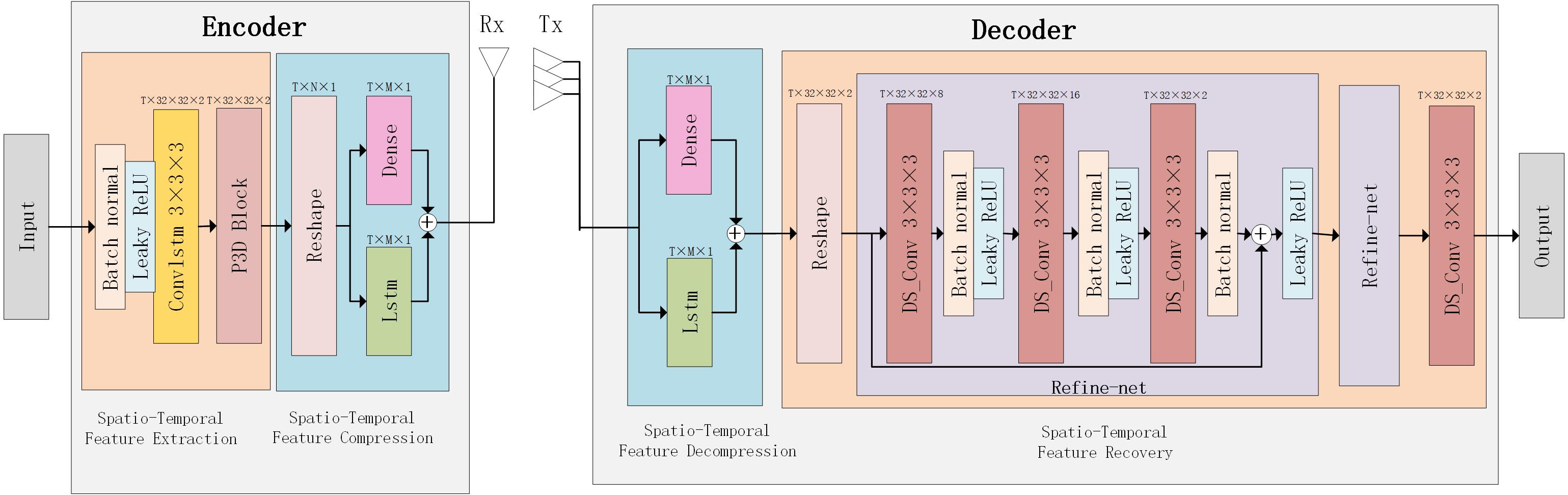}\\
  \caption{The architecture of ConvlstmCsiNet with P3D block}
  \label{archi}
\end{figure*}

\section{CSI Feedback System}

We consider an FDD massive MIMO downlink system with $N_{t}$ transmitting antennas at the BS and a single receiving antenna at each UE, which is operated in OFDM with $\tilde{N_{c}}$ subcarriers. The received signal carried by the $n^{th}$ ($n=1,2,\cdots,\tilde{N_{c}}$) subcarrier can be given as:
\begin{eqnarray}
y_{n}=\tilde{\textbf{h}}_{n}^{H}\textbf{v}_{n}x_{n}+z_{n}
\label{rec_signal}
\end{eqnarray}
where $\tilde{\textbf{h}}_{n}\in \mathbb{C}^{N_{t}}$, ${\textbf{v}}_{n}\in \mathbb{C}^{N_{t}}$, ${x}_{n}\in \mathbb{C}$ and ${z}_{n}\in \mathbb{C}$ denote the instantaneous channel vector, the precoding vector, the modulated transmit symbol and the additional noise at the $n^{th}$ subcarrier, respectively. Then the CSI matrix can be denoted as:
\begin{eqnarray}
\tilde{\textbf{H}}=[\tilde{\textbf{h}}_{1},\tilde{\textbf{h}}_{2},\cdots,\tilde{\textbf{h}}_{\tilde{N_{c}}}]\in \mathbb{C}^{N_{t}\times \tilde{N_{c}}}
\label{csi_matrix}
\end{eqnarray}

We assume that each UE can acquire the estimation of channel response $\tilde{\textbf{H}}$, which is then fed back to the BS to help the BS to generate the precoding vector $\textbf{v}_{n}$. The process of CSI feedback from the UE to the BS, involving the actual required CSI compression, is the main goal of our research. Before being transmitted to the BS, the CSI matrix requires two pretreatments:
\begin{itemize}
\item $\tilde{\textbf{H}}$ is supposed to be sparse in the angular-delay domain after undergoing a 2D discrete Fourier transform (DFT) operation.

\item In the delay domain, most of the elements in $\tilde{\textbf{H}}$ are zeros except for the first few non-zero columns, because the time delay between multipath arrivals around the straight path lies within a small finite time period. Therefore, the first $N_{c}$ ($N_{c}<\tilde{N_{c}}$) nonzero columns can be retained, while the rest are removed, and the new $N_{t}\times N_{c}$ sized CSI matrix is represented as $\textbf{H}$.

\end{itemize}

\par According to \cite{al2007based}, we assume that the channel matrix \textbf{H} remains fixed for a given OFDM symbol and its associated precoding vector, however, it varies from time to time based on a state-space model. Denote that $\textbf{H}_{t}=[\textbf{h}_{1,t},\textbf{h}_{2,t},\cdots,\textbf{h}_{N_{c},t}]\in \mathbb{C}^{N_{t}\times N_{c}}$ is the instantaneous CSI at $t^{th}$ time step, and then $\textbf{H}_{t+1}$ at next time step can be expressed as:
\begin{eqnarray}
\textbf{H}_{t+1}=\textbf{F}\cdot \textbf{H}_{t}+\textbf{G}\cdot \textbf{u}_{t}
\label{ht1}
\end{eqnarray}
where $\textbf{u}_{t}\in \mathbb{C}^{N_{t}\times N_{c}}$ is the additive noise that each element $u_{t}^{(i,j)}\sim N(0, \sigma^{2}_{u})$, and $\textbf{F}, \textbf{G}\in \mathbb{C}^{N_{t}\times N_{t}}$ are the weight square matrices, which are assumed to be available to the receiver. For convenience, we set $\textbf{F}=(1-\alpha^{2})\textbf{I}$ and $\textbf{G}=\alpha^{2} \textbf{I}$ by introducing a new parameter $\alpha$, which depicts the correlation between adjacent CSI matrices. So this sequence of time-varying channel matrix is defined as: $\{\textbf{H}_{t}\}^{T}_{t=1}=\{\textbf{H}_{1},\textbf{H}_{2},\cdots,\textbf{H}_{T}\}$.

\par During transmission, $\{\textbf{H}_{t}\}^{T}_{t=1}$ is separated into a real part and an imaginary part to reduce the computational complexity, where all elements in the matrix are turned into real numbers and normalized within $[0,1]$. With the help of DFT and truncation operations, the number of feedback parameters should be reduced from $\tilde{N}=2\times \tilde{N}_{c}\times N_{t}$ to $N=2\times N_{c}\times N_{t}$, which still remains a large number of parameters in massive MIMO systems and information compression is required during the transmission procedure. The model consists of an encoder at the UE to convert a CSI matrix $\textbf{H}_{t}$ of size $N$ into a compressed $M$-dimensional ($M<N$) codeword $\textbf{s}_{t}$, as well as a decoder at the BS to make the compressed vector $\textbf{s}_{t}$ transform back to the original CSI matrix. The data compression ratio is $\gamma=M/N$. Once the BS completes the recovery of $\textbf{H}_{t}$, i.e., $\hat{\textbf{H}}_{t}$, it outputs the final matrix $\hat{\tilde{\textbf{H}}}_{t}$ by adding zero columns and performing inverse DFT.

\section{Proposed ConvlstmCsiNet with P3D blocks}

The proposed ConvlstmCsiNet is illustrated in Fig.~\ref{archi}. It includes an encoder at the UE and a decoder at the BS. The encoder is divided into two modules, i.e., \emph{feature extraction} and \emph{feature compression};  and the decoder consists of \emph{feature decompression} and \emph{feature recovery} modules, where RefineNet unit is employed in the feature recovery module.

Different types of network layers are colored and each layer has the output shape on the top, marked by $T\times H\times W\times C$ or $T\times L\times C$, where $T$, $H$, $W$, $C$ and $L$ denote the time step of RNN, height, width, channel numbers of feature maps, and codeword length, respectively. After the DFT and truncation operations, the CSI matrix $\textbf{H}$ is then fed into this CSI feedback autoencoder with the input shape of $T\times 32\times 32\times 2$ ($H=N_{t}=32$, $W=N_{c}=32$), where two channels represent the real and imaginary parts of \textbf{H}. The output remains the same shape as the input.

\subsection{ConvlstmCsiNet}
\subsubsection{RNN in Feature Extraction}
 On the basis of CsiNet\cite{ref9}, we refine the feature extraction module by adding a convolutional long short-term memory (ConvLSTM) \cite{xingjian2015convolutional} layer before the convolution, and adopt the memory function of RNN to learn the temporal correlation from the inputs of previous time steps as well as compress the temporal redundancy. Therefore, it can help the convolution to capture more useful temporal information in feature extraction.

\par ConvLSTM is a variant of LSTM, which is proposed in RNN to solve the problem of time sequence gradient disappearing with the increase of calculation time. The main change is that the weight calculation is switched from linear operation to convolution operation, which helps it not only inherit the ability of LSTM and capture the temporal correlation, but also depict the detailed local information in image features like CNN, simultaneously.

\begin{figure}[!ht]
  \centering
  \includegraphics[width=2.0in, trim=0in 0in 0in 0.2in]{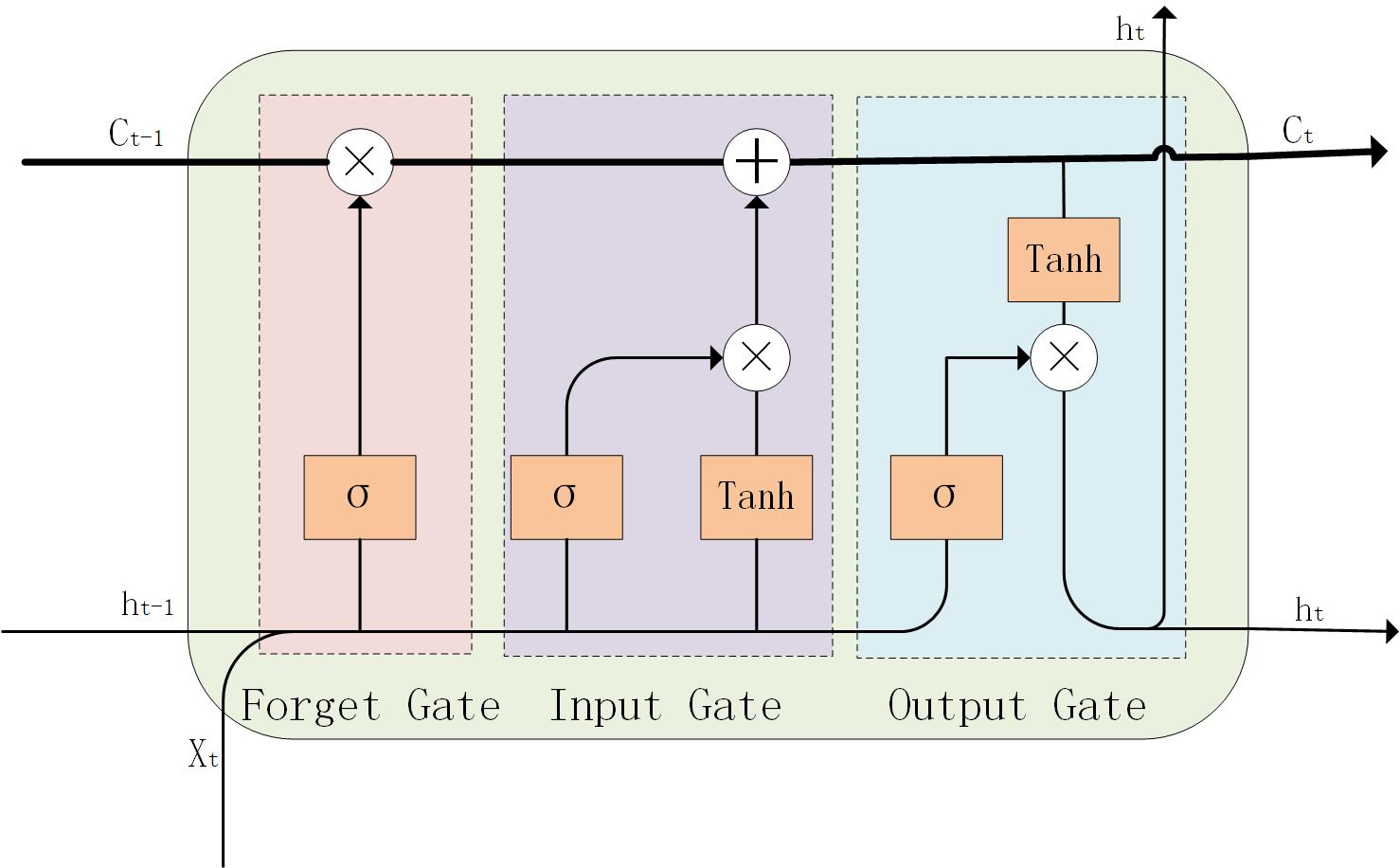}\\
  \caption{The structure of three gates in ConvLSTM \cite{xingjian2015convolutional}}
\label{convlstm}
\end{figure}

The main structure of ConvLSTM is shown in Fig.~\ref{convlstm}. It has the ability to remove or add information to the cell state through three well-designed gates, i.e., forget gate, input gate and output gate, including a sigmoid activation layer and a dot multiplication operation. Since convolution operations require fewer parameters than linear operations, ConvLSTM can help to reduce the size of the model.

\par The symmetric feature compression and feature decompression modules refer to RecCsiNet \cite{lu2018mimo}, which has achieved higher accuracy than PR-RecCsiNet \cite{lu2018mimo} or CsiNet-LSTM \cite{wang2018deep}. It uses two parallel row structures, i.e., the fully-connected (FC) layer and the LSTM layer, to compress the reshaped $N$-length vector into a $M$-length codeword, simultaneously. Then we put the merged codeword as the output of the encoder, and decompress it back to $N$-length with the symmetric structure, which will be reshaped into two $32\times 32$ sized features, serving as a rough estimation of the real and imaginary parts of \textbf{H}. During the feedback transmission, the feedback channel is assumed to be perfect enough to transmit the compressed codeword without any damage or loss.

\par Although ConvLSTM has so many advantages, we retain LSTM instead of completely replacing it with ConvLSTM since LSTM can perform better in terms of overall information interaction due to its FC operation in  weight calculations, thus is more suitable for feature compression, while ConvLSTM is more adaptable for depicting local detailed information.

\subsubsection{Depthwise Separable Convolution in Feature Recovery}

RefineNet in CsiNet \cite{ref9} is adopted as the basic structure. Each RefineNet block has three $3\times3\times3$ Conv3D layers, which are cascaded together one by one, outputting 8, 16 and 2 feature maps, respectively. The feature recovery module helps to refine the primary rough estimation of \textbf{H} with two RefineNet blocks and the results in CsiNet have testified that two blocks are sufficient to recover the CSI matrix and more blocks will lead to parameter redundancy. After two RefineNet blocks follow a $3\times3\times3$ Conv3D layer and a sigmoid activation layer, which outputs the final result of the recovered \textbf{H}, including its real and imaginary parts.
\begin{figure}[!ht]
  \centering
    \includegraphics[width=2.5in, trim=0in 0in 0in 0.3in]{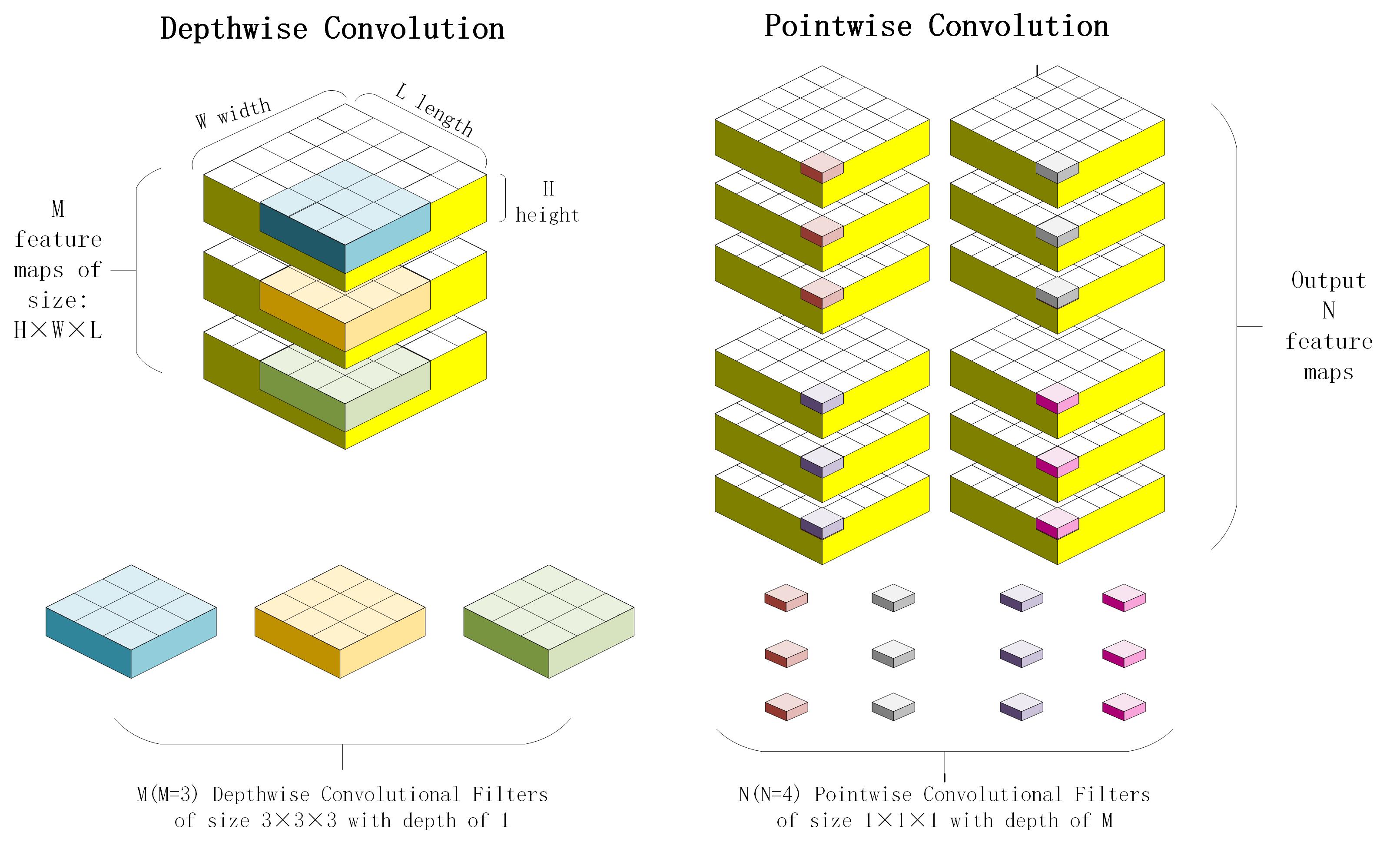}\\
  \caption{Filters of depthwise separable convolution}
  \label{conv}
\end{figure}

While in this module, all standard convolutions in the feature recovery module are replaced by a new type of convolutional layer, i.e., depthwise separable convolution \cite{sandler2018mobilenetv2}, referred to as DS-Conv. This substitution not only reduces the number of parameters, but also helps the RefineNet achieve better performance and higher recovery accuracy. According to MobileNet, it can be divided into two steps: \emph{depthwise convolution} and \emph{pointwise convolution}, the kernels of which are shown in Fig.~\ref{conv}.

It is assumed that the original $3\times3\times3$ Conv3D accepts $M$ input feature maps and outputs $N$ feature maps. Depthwise convolution is a set of convolutions, each of which is responsible for one feature map separately, so there are $M$ $3\times3\times3$ 1-depth depthwise convolution filters to output $M$ feature maps. While pointwise convolution is a $M$-depth $1\times1\times1$ convolution to deal with $M$ feature maps obtained from depthwise convolution and outputs $N$ feature maps. The first step is mainly responsible for capturing features in each channel, while the second step is for the dimensions of ascending and descending channels, as well as for information integration and interaction across channels, which helps the convolution to better understanding the correlation between different channels. The parameter number of DS-Conv3D is $(M\times3^{3}+M\times N)/M\times3^{3}\times N$ time of the Conv3D, so that DS-Conv3D can also help to reduce the parameter size of the feature recovery module to a certain extent. In addition, due to the large use of pointwise convolution, highly optimized matrix multiplications, such as GEMM, can be used directly to complete them without the pre-processing operation of im2col, which greatly improves the operational efficiency.

\subsection{Decoupled Spatial-Temporal Feature Extraction in ConvlstmCsiNet}
\par For further refinement of ConvlstmCsiNet, we focus on the spatial-temporal feature representation in the feature extraction module. Inspired by \cite{qiu2017learning}, we replace the convolutional layer with Pseudo-3D (P3D) in ConvlstmCsiNet to perform feature extraction. The key idea of P3D is to capture features in the temporal and spatial domains, respectively. Suppose we have 3D convolutional filters of size $T_d\times S_d \times S_d$ ($T_d$ and $S_d$ denote temporal depth and spatial depth, respectively), which can be naturally decoupled into $1\times S_d\times S_d$ convolutional filters equivalent to 2D convolutions in spatial domain and $T_d\times1\times1$ convolutional filters equivalent to 1D convolutions on temporal domain. This block replaces the standard convolutional layer with two filters in a cascaded or paralleled manner. In this way, the number of parameters and computational complexity can be reduced.
\begin{figure}[!ht]
  \centering
 \includegraphics[width=2.5in, trim=0in 0in 0in 0.3in]{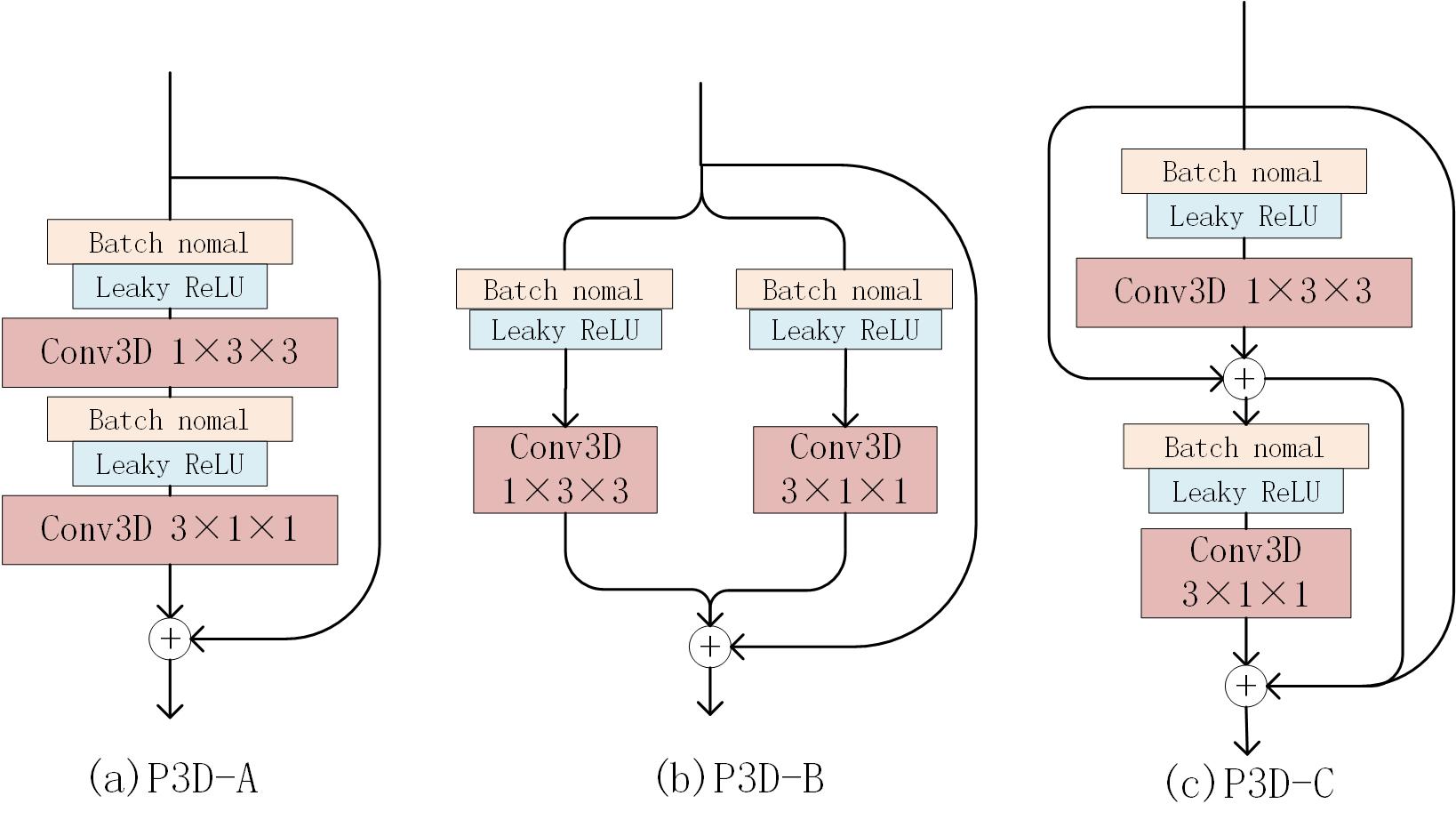}\\
  \caption{Three designs of P3D block}
  \label{P3D}
\end{figure}
%

Considering whether the temporal and spatial filters should directly or indirectly influence each other or the final output, three designs of P3D blocks are proposed, which are shown in Fig.~\ref{P3D}. The skip connection structure of ResNet \cite{resnet} is also used here, which can directly pass the data flow to subsequent layers and lead the model degenerating into a shallow network, thus helping to ease the optimization and improve the robustness of the NN by skipping those unnecessary layers. For regularization, we adopt the idea of pre-activation structure \cite{he2016identity} that the batch normalization (BN) layer followed with an activation layer of leaky ReLU is placed before all weighted layers (e.g. convolutional layer), which has the impacts that the optimization is further eased and the regularization of the model is improved.


\par Based on ConvlstmCsiNet, we name the three newly proposed models as ConvlstmCsiNet-A, ConvlstmCsiNet-B and ConvlstmCsiNet-C, where P3D-A, P3D-B or P3D-C blocks replace the convolution in feature compression module, respectively. Then ConvlstmCsiNet is used to highlight the effect of P3D in feature extraction by comparing to those with P3D blocks.
Define the network as an autoencoder function of the input $\textbf{H}_{t}$, then the output can be expressed as:
\begin{eqnarray}
\begin{aligned}
\hat{\textbf{H}}_{t}&:= f(\{\textbf{H}_{k}\}^{t}_{k=1};\Theta) \\
&=f_{dec}(f_{enc}(\{\textbf{H}_{k}\}^{t}_{k=1};\Theta_{enc});\Theta_{dec})
\end{aligned}
\end{eqnarray}
where $\Theta$ is the whole parameters and $f(\cdot )$ represents the function of the network. $f_{dec}$, $f_{enc}$, $\Theta_{dec}$ and $\Theta_{enc}$ denote the maps and parameters of the decoder and encoder, respectively.

\par All networks are trained end-to-end by updating parameters in the procedure of minimizing the mean squared error (MSE) loss function using the ADAM algorithm, which can be given as follows:
\begin{eqnarray}
\begin{aligned}
L(\Theta)&=\dfrac{1}{MT}\sum^{M}_{m=1}\sum^{T}_{t=1}\|f(\textbf{H}_{m,t};\Theta)- \textbf{H}_{m,t}\|_{2}^{2}\\
&= \dfrac{1}{MT}\sum^{M}_{m=1}\sum^{T}_{t=1}\sum^{N_{t}}_{i=1}\sum^{N_{c}}_{j=1}|f(\textbf{H}_{m,t}^{(i,j)};\Theta)- \textbf{H}_{m,t}^{(i,j)}|
\end{aligned}
\end{eqnarray}
where $\|\cdot \|_{2}$ is the Euclidean norm, $T$ and $M$ denote the number of recurrent steps and the total number of examples in the training data, respectively.

\section{Experiments and Numeral Results}
In this section, we illustrate the training process in details and discuss the experimental results compared with several other methods of CSI feedback compression networks.

\par Two metrics are introduced to evaluate the models:

\begin{itemize}
\item \textbf{Normalized Mean Square Error (NMSE)}: it quantifies the difference between the input $\{\textbf{H}_{t}\}^{T}_{t=1}$ and the output $\{\hat{\textbf{H}}_{t}\}^{T}_{t=1}$, which can be defined as:
\begin{eqnarray}
\mathrm{NMSE}=\mathbb{E}\bigg\{\dfrac{1}{T}\sum^{T}_{t=1} \dfrac{\|\textbf{H}_{t}-\hat{\textbf{H}}_{t}\|_{2}^{2}}{\|\textbf{H}_{t}\|_{2}^{2}}\bigg\}
\end{eqnarray}
\item \textbf{Cosine Similarity $\rho$}: it depicts the similarity between the original CSI matrix $\tilde{\textbf{H}}$ and the recovered $\hat{\tilde{\textbf{H}}}$ by calculating cosine similarity within the
channel response $\tilde{\textbf{h}}_{n, t}$ ($n=1,\cdots,\tilde{N}_{c}$) of each subcarrier, which is given as:
\begin{eqnarray}
\rho=\mathbb{E}\bigg\{\frac{1}{T}\frac{1}{\tilde{N}_{c}}\sum^{T}_{t=1}\sum^{\tilde{N}_{c}}_{n=1}\frac{|\hat{\tilde{\textbf{h}}}_{n,t}^{H}\cdot \tilde{\textbf{h}}_{n,t}|}{\|\hat{\tilde{\textbf{h}}}_{n,t}\|_{2} \|\tilde{\textbf{h}}_{n,t}\|_{2}}\bigg\}
\end{eqnarray}
\end{itemize}

\par The MIMO-OFDM feedback system is set to work with $\tilde{N}_{c}=1,024$ subcarriers and uniform linear array (ULA) with $N_{t}=32$ antennas at the BS. After the DFT and truncation operations, only the first $N_{c}=32$ columns in CSI feedback matrix $\mathbf{H}$ are nonzero and remain unchanged, which turns $\mathbf{H}$ from a $1,024\times32$ shape to a new $32\times32$ shape.  According to Eq.~\ref{ht1}, we add tiny white Gauss noise ($\sigma_{u}=10^{-3}$) and coloration index $\alpha$ between each time step, and the 2D CSI feedback matrix can be extended to a $T$-time sequence of time-varying CSI matrix, where $T$ is the recurrent time steps and is set to 4 for convenience.

All examples of $\mathbf{H}$ are generated based on the COST2100 \cite{liu2012cost} channel model. We use the indoor picocellular scenario at the 5.3 GHz band, and all parameters follow the default setting in \cite{liu2012cost}. During the training process of each model, we use 100,000 examples for training, 30,000 for validation and 20,000 for testing. The learning rate is set to $10^{-3}$ for the first 1,000 epochs, $5\times10^{-4}$ for the middle $1,000-1,200$ epochs and $10^{-4}$ for the last $1,200-1,500$ epochs.


\renewcommand\arraystretch{1.04}
\begin{table}[!ht]
\footnotesize
\centering
\caption{NMSE and $\rho$ in different CRs when $\alpha=0.1$}
\begin{tabular}{|l|l|llll|}
\hline
\multicolumn{6}{|c|}{Indoor Scenario}                                                                                                                                  \\ \cline{2-6}
\hline
    &CR                            & \multicolumn{1}{l|}{1/4} & \multicolumn{1}{l|}{1/8} & \multicolumn{1}{l|}{1/16} & \multicolumn{1}{l|}{1/32} \\ \hline
\multicolumn{1}{|c|}{\multirow{6}{*}{\rotatebox{90}{NMSE}}}
                                            & CsiNet                & -17.5                    & -12.3                    & -9.93                     & -6.98                   \\ \cline{2-2}
\multicolumn{1}{|c|}{}                      & RecCsiNet             & -21.5                    & -18.8                    & -16.8                     & -13.4                         \\ \cline{2-2}
\multicolumn{1}{|c|}{}                      & ConvlstmCsiNet-A      & \textbf{-28.4}           & \textbf{-23.5}           & \textbf{-20.7}            & \textbf{-15.0}                         \\ \cline{2-2}
\multicolumn{1}{|c|}{}                      & ConvlstmCsiNet-B      & -25.9                    & -20.7                    & -18.3                     & -14.0                         \\ \cline{2-2}
\multicolumn{1}{|c|}{}                      & ConvlstmCsiNet-C      & -26.5                    & -22.0                    & -19.0                     & -14.4                         \\ \cline{2-2}
\multicolumn{1}{|c|}{}                      & ConvlstmCsiNet        & -24.9                    & -23.0                    & -18.7                     & -13.5                         \\ \cline{1-2}
\hline

\multicolumn{1}{|c|}{\multirow{6}{*}{\rotatebox{90}{$\rho$}}}       & CsiNet                      & 95.1$\%$                    & 93.1$\%$                    & 90.4$\%$                     & 87.4$\%$                         \\ \cline{2-2}
\multicolumn{1}{|c|}{}                      & RecCsiNet             & 95.7$\%$                    & 95.0$\%$                    & 94.7$\%$                    & 93.3$\%$                         \\ \cline{2-2}
\multicolumn{1}{|c|}{}                      & ConvlstmCsiNet-A      & \textbf{95.8$\%$}           & \textbf{95.7$\%$ }          & \textbf{95.5$\%$ }          & \textbf{94.2$\%$ }                         \\ \cline{2-2}
\multicolumn{1}{|c|}{}                      & ConvlstmCsiNet-B      & \textbf{95.8$\%$}           & 95.4$\%$                    & 95.0$\%$                    & 93.8$\%$                         \\ \cline{2-2}
\multicolumn{1}{|c|}{}                      & ConvlstmCsiNet-C      & \textbf{95.8$\%$  }         & 95.6$\%$                    & 95.3$\%$                    & 93.7$\%$                         \\ \cline{2-2}
\multicolumn{1}{|c|}{}                      & ConvlstmCsiNet        & 95.7$\%$                    & \textbf{95.7$\%$}           & 95.2$\%$                    & 93.5$\%$                         \\ \cline{1-2}
\hline
\end{tabular}
\label{NMSE}
\end{table}

\par Since the DL-based approaches are superior to the traditional CS-based methods, we only compare our methods with the DL-based approaches, such as CsiNet \cite{ref9} and RecCsiNet \cite{lu2018mimo}). The corresponding NMSE and $\rho$ of each network are given in Table~\ref{NMSE}, where the best results are marked in bold. The value of NMSE is too small that we use log(NMSE) to represent it. Obviously, our proposed model ConvlstmCsiNet-A can achieve the best performance on both NMSE and $\rho$.

\renewcommand\arraystretch{1.09}
\begin{table}[!ht]
\footnotesize
\centering
\caption{Percentage improvement of proposed networks compared with CsiNet $\&$ RecCsiNet}
\begin{tabular}{|l|l|l|llll|}
\hline
\multicolumn{3}{|c|}{CR}                              & \multicolumn{1}{l|}{1/4} & \multicolumn{1}{l|}{1/8} & \multicolumn{1}{l|}{1/16} & \multicolumn{1}{l|}{1/32} \\ \hline
\multirow{8}{*}{\rotatebox{90}{Compare to CsiNet}}    & \multirow{4}{*}{\rotatebox{90}{NMSE}}
         & ConvlstmCsiNet-A         & 64.0$\%$   & 91.1$\%$   & 108.5$\%$      & 114.9$\%$           \\ \cline{3-3}
&        & ConvlstmCsiNet-B         & 48.0$\%$   & 68.3$\%$   & 84.3$\%$       & 100.6$\%$           \\ \cline{3-3}
&        & ConvlstmCsiNet-C         & 51.4$\%$   & 78.9$\%$   & 91.3$\%$       & 106.3$\%$           \\ \cline{3-3}
&        & ConvlstmCsiNet           & 42.3$\%$   & 87.0$\%$   & 88.3$\%$       & 93.4$\%$           \\ \cline{2-3}
                                                      & \multirow{4}{*}{\rotatebox{90}{$\rho$}}
         & ConvlstmCsiNet-A         & 0.73$\%$    & 2.8$\%$    & 5.6$\%$         & 7.8$\%$                          \\ \cline{3-3}
&        & ConvlstmCsiNet-B         & 0.73$\%$    & 2.5$\%$    & 5.1$\%$         & 7.3$\%$                          \\ \cline{3-3}
&        & ConvlstmCsiNet-C         & 0.73$\%$    & 2.7$\%$    & 5.4$\%$         & 7.2$\%$                          \\ \cline{3-3}
&        & ConvlstmCsiNet           & 0.63$\%$    & 2.8$\%$    & 5.3$\%$         & 7.0$\%$                          \\ \cline{1-3}
\hline
\multirow{8}{*}{\rotatebox{90}{Compare to RecCsiNet}}  & \multirow{4}{*}{\rotatebox{90}{NMSE}}
         & ConvlstmCsiNet-A         & 32.1$\%$   & 25.0$\%$   & 23.2$\%$       & 11.9$\%$                          \\ \cline{3-3}
&        & ConvlstmCsiNet-B         & 20.5$\%$   & 10.1$\%$   & 8.93$\%$       & 4.48$\%$                          \\ \cline{3-3}
&        & ConvlstmCsiNet-C         & 23.3$\%$   & 17.0$\%$   & 13.1$\%$       & 7.46$\%$                          \\ \cline{3-3}
&        & ConvlstmCsiNet           & 15.8$\%$   & 22.3$\%$   & 11.3$\%$       & 0.75$\%$                          \\ \cline{2-3}
                                      & \multirow{4}{*}{\rotatebox{90}{$\rho$}}
         & ConvlstmCsiNet-A         & 0.10$\%$       & 0.74$\%$        & 0.84$\%$         & 0.96$\%$                          \\ \cline{3-3}
&        & ConvlstmCsiNet-B         & 0.10$\%$       & 0.42$\%$        & 0.32$\%$         & 0.54$\%$                          \\ \cline{3-3}
&        & ConvlstmCsiNet-C         & 0.10$\%$       & 0.63$\%$        & 0.63$\%$         & 0.43$\%$                          \\ \cline{3-3}
&        & ConvlstmCsiNet           & 0.00$\%$       & 0.74$\%$        & 0.53$\%$         & 0.21$\%$                          \\ \cline{1-3}
\hline
\end{tabular}
\label{improve}
\end{table}
To show the contrast more intuitively, we give percentage improvements of the proposed network compared with CsiNet and RecCsiNet in Tabel \ref{improve}. It demonstrates that all four purposed models outperform CsiNet and RecCsiNet.
 In the networks with P3D blocks, ConvlstmCsiNet-A achieves the best performance while ConvlstmCsiNet-B achieves the worst, indicating that the cascaded manner of temporal and spatial filter performs better than the parallel fashion, which can also be proved by the result that the performance of the combined structure ConvlstmCsiNet-C is between ConvlstmCsiNet-A and ConvlstmCsiNet-B.

When analyzing the functions of P3D blocks, all ConvlstmCsiNet-A, ConvlstmCsiNet-B and ConvlstmCsiNet-C have obtained much lower NMSE and higher cosine similarity $\rho$ than ConvlstmCsiNet, especially at high CRs, indicating that the decoupling convolution structure (P3D block) does have a positive impact on capturing features and improving the performance of the network.

\par In Table~\ref{improve}, we can find that in the first part (compared with CsiNet) that the improvements of all four networks are increasing as CR decreases due to a better and more complicated devised architecture. However, the increase in improvement becomes slower when compared with RecCsiNet, which indicates that the advantage of the ConvLSTM layer in feature extraction module in our models becomes less noticeable compared with RecCsiNet at low CRs. This is because the CR value only affects the performance of feature compression and decompression, where LSTM begins to play a major role in accelerating the convergence of models, emphasizing the benefits of LSTM and shrinking the advantage effects of feature extraction part.
\begin{figure}[!ht]
  \centering
 \includegraphics[width=3.1in]{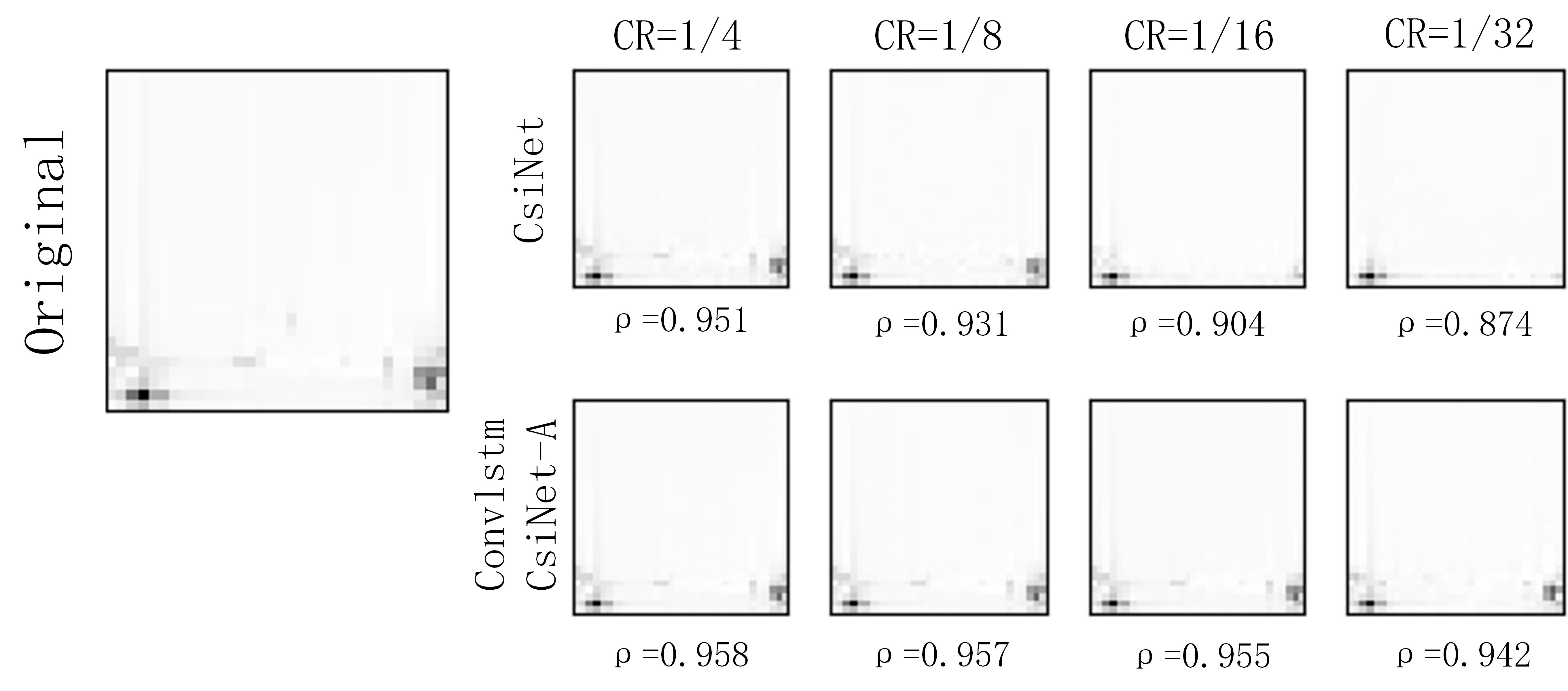}\\
  \caption{Absolute value of original ($\alpha=0.1$) and reconstructed CSI images at different CRs}
\label{contrast}
\end{figure}

Figure~\ref{contrast} plots the reconstructed CSI images by CsiNet and ConvlstmCsiNet-A (the best model we propose) in Pseudo-gray. Obviously, ConvlstmCsiNet-A outperforms CsiNet, especially at low CRs. Moreover, CsiNet may lose some feature information while ConvlstmCsiNet-A does not.
\begin{figure}[!ht]
  \centering
  \includegraphics[width=3.0in]{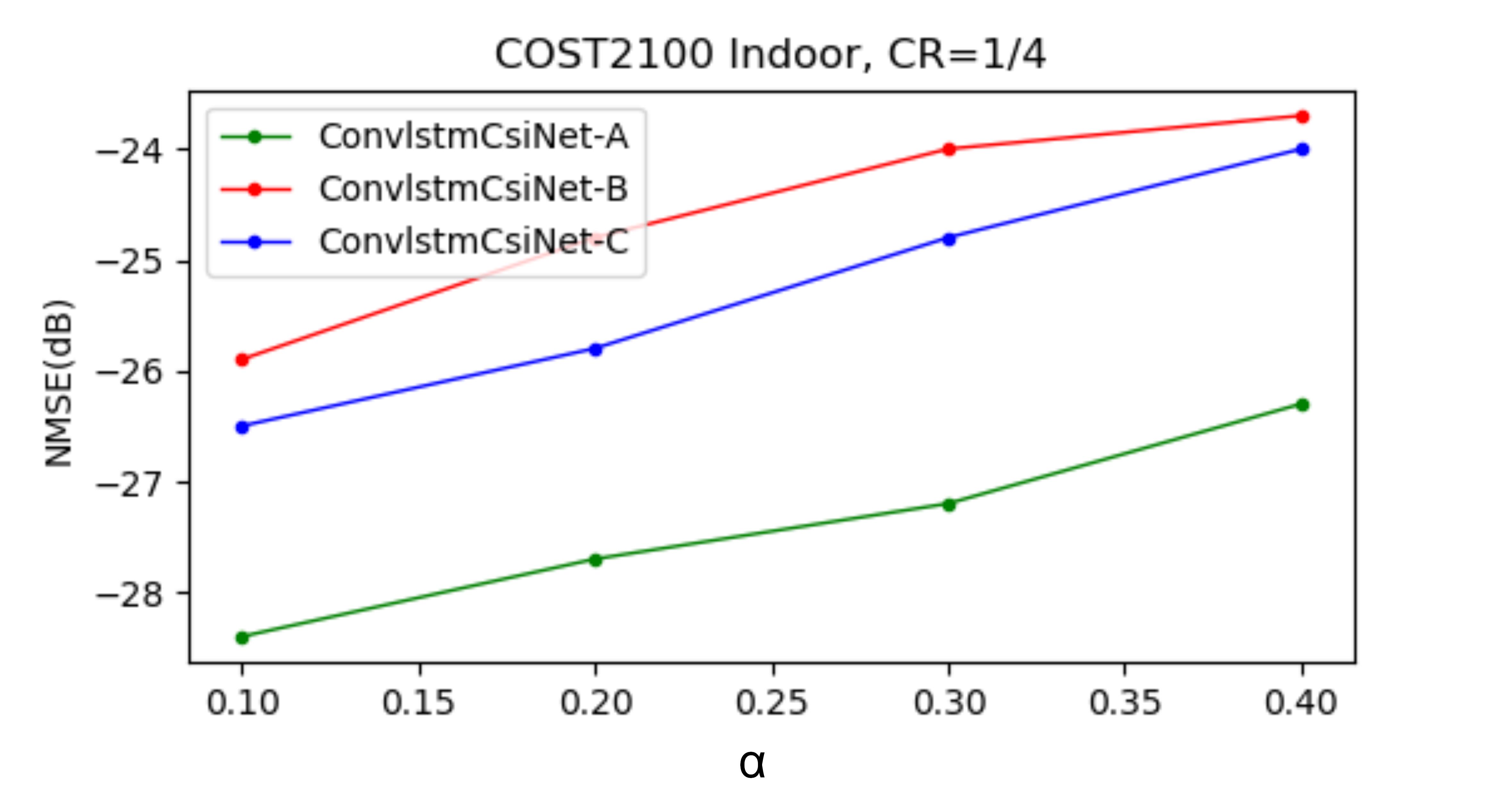}\\
  \caption{NMSE of the proposed NN at CR=1/4 in different correlation parameter $\alpha$}
\label{correlation}
\end{figure}

Figure~\ref{correlation} demonstrates that the rise of $\alpha$ leads to a growth of corresponding NMSE, indicating that a decrease in temporal correlation may prevent the proposed networks from achieving high performance in CSI recovery.

\section{Conclusion}
We proposed a novel network architecture of CSI feedback by adopting RNN and depthwise separable convolution in feature extraction and recovery modules, respectively. Furthermore, we also devised the feature extraction part by studying the decoupled temporal-spacial convolutional representations, which proved to be better than standard Conv3D convolutions. Experimental results demonstrate that our method can improve the performance of RecCsiNet in terms of recovery robustness, accuracy and quality. This architecture has the potential for practical deployment on real MIMO systems.


\bibliographystyle{ieeetr}
\bibliography{bibref}
\end{document}